\documentclass[aps,prb,showpacs,twocolumn,floats]{revtex4}
\usepackage{amssymb}

\usepackage{amsmath}
\usepackage{graphicx,psfig}
\usepackage{dcolumn}
\usepackage{bm}

\input{tcilatex}

\begin{document}

\title{Extended Debye Model for Molecular Magnets}
\author{D. A. Garanin}
\affiliation{ \mbox{${}^1$Department of Physics and Astronony, Lehman College, City
University of New York,} \\ \mbox{250 Bedford Park Boulevard
West, Bronx, New York 10468-1589, U.S.A.} }
\date{\today}

\begin{abstract}
Heat capacity data on Mn$_{12}$ are fitted within the extended Debye model
that takes into account a continuum of optical modes as well as three
different speeds of sound.
\end{abstract}
\maketitle


Molecular magnets (MM) such as Mn$_{12}$ (Ref.\ \onlinecite{lis80}) are
relatively new materials that have a giant effective molecular spin $S$
(such as $S=10$ for Mn$_{12})$ built from several atomic spins by a strong
intramolecular exchange interaction. The magnetic molecules have a uniaxial
anisotropy that is responsible for magnetic bistability and long relaxation
over the barrier at low temperatures.\cite{sesgatcannov93nat} As the
magnetic core of these molecules is surrounded by organic ligands, the
exchange interaction between different molecules building a crystal lattice
is very small. This allows them to relax independently from each other, in
contrast to ferromagnets. A fascinating phenomenon discovered in molecular
magnets is resonance spin tunneling under the barrier that happens if the
energy levels of the spin $S$ in both wells match.\cite
{frisartejzio96prl,heretal96epl,thoetal96nat}

Molecular magnets is a new type of condensed magnetic systems whose
properties differ from those of ferromagnets and dilute paramagnets.
Although in the most temperature range MM are paramagnetic, their relaxation
can differ from that of a single spin embedded in an elastic matrix. Since
the wave length of emitted and absorbed phonons or photons exceeds the
lattice spacing, there can be pronounced coherence effects in relaxation
such as superradiance. \cite{dic54} Photon\cite{chugar02prl} and phonon\cite
{chugar04prl} superradiance in MM can increase relaxation rates by a huge
factor. On the other hand, the opposite effect for initial states of spins
with random phases should lead to suppression of the rates by a huge factor.
Strong inhomogeneous broadening in MM tends to destroy coherence effects,
however, so that efforts should be done to understand the relaxation data.
Another collective phenomenon in relaxation that is not yet fully understood
theoretically is the phonon bottleneck (see Refs.\
\onlinecite
{rubbenjef62pr,abrble70} for older reference and Refs.\
\onlinecite
{gar07prb,gar08prb} for recent work).

To be able to test more sophisticated collective models of relaxation in MM,
one should have reliable theoretical estimations of the single-spin
relaxation rates, most notably the one-phonon or direct relaxation rate. The
latter can depend, in general, on spin-phonon couplings that are difficult
to measure. On the other hand, there is a simple mechanism of spin-lattice
coupling through rotations of the magnetic molecules by transverse phonons
\cite{chu04prl,chugarsch05prb} that can serve at least as the low bound on
spin-lattice relaxation. As in this mechanism the crystal field acting on
the spin is not distorted but only rotated, no unknown coupling constants
enter the theory. Also this mechanism is likely to be the dominating
relaxation channel since the cores of magnetic molecules\ should be much
less deformable than the ligands. The corresponding results for the
relaxation rates $\Gamma $ due to direct processes, as well as the Raman
processes, depend on only one parameter that is currently not precisely
known, the speed of transverse sound $v_{t}$. For direct processes one has $%
\Gamma \varpropto 1/v_{t}^{5},$ whereas for Raman processes $\Gamma
\varpropto 1/v_{t}^{10}.$ Thus the uncertainties of $v_{t}$ dramatically
amplify in the relaxation rates.

In the absence of direct measurements of the speed of sound, the latter can
be extracted from the heat capacity of the lattice by fitting the measured $%
C(T)$ to the Debye theory and extracting the Debye temperature $\Theta _{%
\mathrm{D}}$ that is proportional to the speed of sound. So $\Theta _{%
\mathrm{D}}=38$ K of Ref.\ \onlinecite{gometal98prb} results in $v=1600$
m/s, whereas $\Theta _{\mathrm{D}}=41$ K of Ref.\ \onlinecite{fometal99prb}
results in $v=1727$ m/s. In fact, determination of $\Theta _{\mathrm{D}}$ in
Refs.\ \onlinecite
{gometal98prb,fometal99prb} relies on the low-temperature data, where the
phonon contribution to the heat capacity is $C_{\mathrm{ph}}\varpropto T^{3}$
and the Debye model with the rigid the cut-off at the Brillouin-zone
boundary, that is a crude approximation, is not actually used.

A problem with the extracting the Debye temperature and the speed of sound
above is the assumption that the three branches of acoustic phonons in the
crystal have the same speed $v.$ Indeed, at low temperatures one has $C_{%
\mathrm{ph}}\varpropto \left( v_{1}^{-3}+v_{2}^{-3}+v_{3}^{-3}\right) T^{3}$%
, and there is no way to find $v_{i}$ separately from this formula. On the
other hand, acoustic phonon modes in such a complicated crystal as Mn$_{12}$
should not be strictly longitudinal and strictly transverse and all three
speeds of sound, as well as all three Debye temperatures, should be
different. If they all differ much, the smallest of them dominates the
low-temperature heat capacity and, to a much greater extent, the spin-phonon
relaxation rates. If one assumes that there are two degenerate transverse
phonon modes with $v_{1}=v_{2}=v_{t}\ll v_{3}\equiv v_{l},$ then instead of $%
\Theta _{\mathrm{D}}=38$ K and $v=1977$ m/s one obtains $\Theta _{\mathrm{D,}%
t}=33$ K and $v_{t}=1717$ m/s. If there is one phonon mode with $v_{1}\ll
v_{2,3},$ then one obtains $\Theta _{\mathrm{D,}1}=26.3$ K and $v_{1}=1368$
m/s.

To improve the description and extract more information, one can use the
heat-capacity data at higher temperatures where the contributions from the
phonon modes add up in a different way. This is where the Debye model begins
to really work and crudeness is introduced into the theory. In addition,
optical modes become very important and should be taken into account. In
recent Ref.\ \onlinecite{evaluimetsesjon05prl} on another MM Fe$_{8}$ the
Debye model has been extended by adding an Einstein oscillator that accounts
for all optical modes, and the analysis in a broader temperature range
rendered $\Theta _{\mathrm{D}}=19$ K (as well as the Einstein temperature $%
\Theta _{\mathrm{E}}=38$ K) that differs much from the previousely extracted
value $\Theta _{\mathrm{D}}=34$ K for this MM$.$ On the other hand, all
three speeds of sound were considered as the same in this analysis.

\begin{figure}[t]
\unitlength1cm
\begin{picture}(11,5.5)
\centerline{\psfig{file=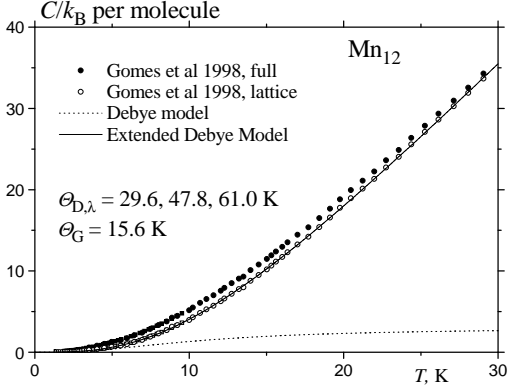,angle=-90,width=8cm}}
\end{picture}
\caption{Heat capacity of Mn$_{12}.$ The standard Debye model fails
everywhere except for very low temperatures, while the extended Debye model
perfectly fits the data.}
\label{Fig-C}
\end{figure}

The data of Ref.\ \onlinecite{gometal98prb} shown in Fig.\ \ref{Fig-C} go as
$C\varpropto T$ at elevated temperatures, and the heat capacity per molecule
by far exceeds the value $3k_{\mathrm{B}}$ of a crystal with one atom per
unit cell at $T\gtrsim \Theta _{\mathrm{D}}$. This means that there are a
lot of optical modes forming a continuum with a nearly constant density of
states. This is expected for molecular magnets having hundreds of atoms
within the unit cell. Description based on a continuum of optical modes is
much more reasonable than the Einstein theory with all due respect. The aim
of the present paper is thus to formulate the extended Debye model (EDM)
including a continuum of optical modes. This will be used to extract the
speeds of sound in Mn$_{12}$ from the experimental data without making the
assumption $v_{1}=v_{2}=v_{3}.$

The thermal energy of acoustic phonons per unit cell of a crystal is given
by
\begin{equation}
U_{\mathrm{ph}}=\frac{1}{N}\sum_{\mathbf{k}\lambda }\frac{\hbar \omega _{%
\mathbf{k}\lambda }}{\exp \left( \frac{\hbar \omega _{\mathbf{k}\lambda }}{%
k_{\mathrm{B}}T}\right) -1},  \label{UphDef}
\end{equation}
where $N$ is the number of unit cells in the crystal, $\omega _{\mathbf{k}%
\lambda }$ are phonon frequencies and $\lambda =1,2,3$ are phonon
polarizations. One can replace summation over $\mathbf{k}$ by integration.
Within the Debye model one assumes that the relation $\omega _{\mathbf{k}%
\lambda }=v_{\lambda }k$ holds everywhere in the Brillouin zone that is
approximated by a sphere bound by the Debye wave vector $k_{\mathrm{D}}.$
The latter is defined by the requirement that that total number of phonon
modes is $N:$%
\begin{equation}
1=\frac{1}{N}\sum_{\mathbf{k}}=v_{0}\int_{0}^{k_{D}}\frac{4\pi k^{2}dk}{%
\left( 2\pi \right) ^{3}}=v_{0}\frac{k_{\mathrm{D}}^{3}}{6\pi ^{2}},
\label{Normalization}
\end{equation}
where $v_{0}$ is the unit-cell volume. This yields
\begin{equation}
k_{\mathrm{D}}=\left( 6\pi ^{2}/v_{0}\right) ^{1/3}.  \label{kDDef}
\end{equation}
One can introduce Debye frequencies $\Omega _{\mathrm{D},\lambda }$ and
Debye temperatures $\Theta _{\mathrm{D},\lambda }$ for different acoustic
phonon branches $\lambda $ as
\begin{equation}
\Omega _{\mathrm{D},\lambda }=v_{\lambda }k_{\mathrm{D}},\qquad k_{\mathrm{B}%
}\Theta _{\mathrm{D},\lambda }=\hbar \Omega _{\mathrm{D},\lambda }.
\label{OmegaDDef}
\end{equation}
Now Eq.\ (\ref{UphDef}) can be written as
\begin{equation}
U_{\mathrm{ph}}=\sum_{\lambda }\int d\omega \rho _{\lambda }(\omega )\frac{%
\hbar \omega }{\exp \left( \frac{\hbar \omega }{k_{\mathrm{B}}T}\right) -1},
\label{Uphviarho}
\end{equation}
where the densities of states are given by
\begin{equation}
\rho _{\lambda }(\omega )=3\omega ^{2}/\Omega _{\mathrm{D},\lambda }^{3}
\label{rhoAcou}
\end{equation}
for $\omega <\Omega _{\mathrm{D},\lambda }$ and zero otherwise. At low
temperatures, $T\ll \Theta _{\mathrm{D}},$ Eq.\ (\ref{Uphviarho}) yields
\begin{equation}
U_{\mathrm{ph}}=k_{\mathrm{B}}\frac{\pi ^{4}}{5}\sum_{\lambda }\frac{T^{4}}{%
\Theta _{\mathrm{D},\lambda }^{3}}.  \label{UphLT}
\end{equation}
and thus
\begin{equation}
C_{\mathrm{ph}}=\frac{dU_{\mathrm{ph}}}{dT}=k_{\mathrm{B}}\frac{4\pi ^{4}}{5}%
\sum_{\lambda }\frac{T^{3}}{\Theta _{\mathrm{D},\lambda }^{3}}.
\label{CphLT}
\end{equation}
At high temperatures, $T\gg \Theta _{\mathrm{D}},$ Eq.\ (\ref{Uphviarho})
yields
\begin{equation}
U_{\mathrm{ph}}=3k_{\mathrm{B}}T,\qquad C_{\mathrm{ph}}=3k_{\mathrm{B}}.
\label{UC-HT}
\end{equation}
For all $\Theta _{\mathrm{D},\lambda }$ being the same, the coefficient in
Eq.\ (\ref{CphLT}) is a huge number $12\pi ^{4}/5\simeq 234.$ Because of
this, Eq.\ (\ref{CphLT}) does not smoothly join with Eq.\ (\ref{UC-HT}) at $%
T\sim \Theta _{\mathrm{D}}$, and the applicability of Eqs.\ (\ref{UphLT})
and (\ref{CphLT}) requires in fact very low temperatures, not just $T\ll
\Theta _{\mathrm{D}}.$ On the other hand, Eq.\ (\ref{UC-HT}) is at striking
contradiction with the experiments shown in Fig.\ \ref{Fig-C} because of the
huge unaccounted contribution of optical modes. Thus the usefulness of the
Debye model in its standard form is limited, at least for molecular magnets.

To improve the Debye model in a minimal way, one can add optical modes with
a constant density of states
\begin{equation}
\rho _{\lambda }(\omega )=3/\Omega _{\mathrm{G}},\qquad \omega \geq \Omega _{%
\mathrm{D},\lambda }  \label{OmegaGDef}
\end{equation}
where $\Omega _{\mathrm{G}}$ is another characteristic frequency that should
be considered as a fitting parameter. With Eqs.\ (\ref{rhoAcou}) and (\ref
{OmegaGDef}) inserted, Eq.\ (\ref{Uphviarho}) becomes
\begin{eqnarray}
\frac{U_{\mathrm{ph}}}{k_{\mathrm{B}}} &=&3\sum_{\lambda }\frac{T^{4}}{%
\Theta _{\mathrm{D},\lambda }^{3}}F_{3}\left( \frac{\Theta _{\mathrm{D}%
,\lambda }}{T}\right)  \notag \\
&&-\frac{3T^{2}}{\Theta _{\mathrm{G}}}\sum_{\lambda }F_{1}\left( \frac{%
\Theta _{\mathrm{D},\lambda }}{T}\right) +\frac{3}{2}\pi ^{2}\frac{T^{2}}{%
\Theta _{\mathrm{G}}},  \label{UphEDM}
\end{eqnarray}
where $k_{\mathrm{B}}\Theta _{\mathrm{G}}=\hbar \Omega _{\mathrm{G}}$ and
\begin{equation}
F_{n}\left( y\right) \equiv \int_{0}^{y}dx\frac{x^{n}}{e^{x}-1}.
\label{FnDef}
\end{equation}
For the heat capacity one obtains
\begin{eqnarray}
\frac{C_{\mathrm{ph}}}{k_{\mathrm{B}}} &=&12\sum_{\lambda }\frac{T^{3}}{%
\Theta _{\mathrm{D},\lambda }^{3}}F_{3}\left( \frac{\Theta _{\mathrm{D}%
,\lambda }}{T}\right) -\frac{6T}{\Theta _{\mathrm{G}}}\sum_{\lambda
}F_{1}\left( \frac{\Theta _{\mathrm{D},\lambda }}{T}\right)  \notag \\
&&+3\sum_{\lambda }\left( \frac{\Theta _{\mathrm{D},\lambda }}{\Theta _{%
\mathrm{G}}}-1\right) \frac{\Theta _{\mathrm{D},\lambda }/T}{e^{\Theta _{%
\mathrm{D},\lambda }/T}-1}+3\pi ^{2}\frac{T}{\Theta _{\mathrm{G}}}.
\label{CphEDM}
\end{eqnarray}
In the high-temperature limit $T\gg \Theta _{\mathrm{D}}$ these equations
yield
\begin{equation}
U_{\mathrm{ph}}=\frac{3}{2}\pi ^{2}k_{\mathrm{B}}\frac{T^{2}}{\Theta _{%
\mathrm{G}}},\qquad C_{\mathrm{ph}}=3\pi ^{2}k_{\mathrm{B}}\frac{T}{\Theta _{%
\mathrm{G}}},  \label{UCEDMHT}
\end{equation}
instead of Eq.\ (\ref{UC-HT}) and in accord with the experimental data of
Ref.\ \onlinecite{gometal98prb} shown in Fig.\ \ref{Fig-C}.

\begin{figure}[t]
\unitlength1cm
\begin{picture}(11,5.5)
\centerline{\psfig{file=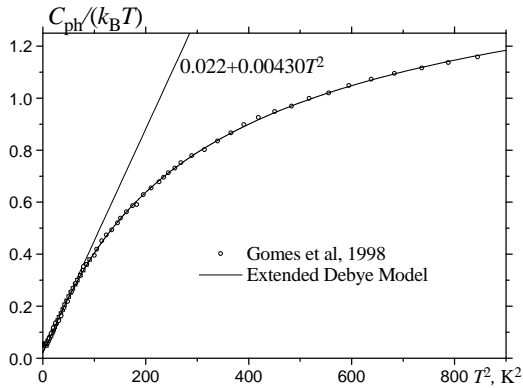,angle=-90,width=8cm}}
\end{picture}
\caption{Heat capacity of Mn$_{12}$ in $C(T)/T$ vs $T^{2}$ representation
used to eliminate the small parasite term $C(T)\varpropto T$.}
\label{Fig-C_over_T}
\end{figure}

\begin{figure}[t]
\unitlength1cm
\begin{picture}(11,5.5)
\centerline{\psfig{file=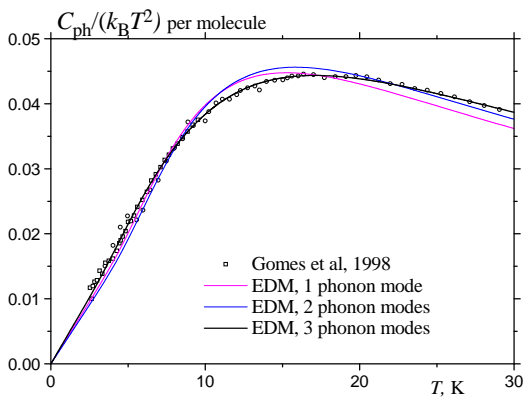,angle=-90,width=8cm}}
\end{picture}
\caption{Heat capacity of Mn$_{12}$ in the presentation used as a crucial
test for different models. One can see that only the model with three
different acoustic phonon modes fits the data.}
\label{Fig-C_over_T2}
\end{figure}

To fit the experimental heat capacity with Eq.\ (\ref{CphEDM}), one has
first to subtract the spin (Schottky) contribution $C_{S}(T)$ from $C(T).$
Using the spin Hamiltonian in zero field
\begin{equation}
\hat{H}=-DS_{z}^{2}-AS_{z}^{4}+\hat{H}^{\prime },  \label{HamMn12}
\end{equation}
where $D/k_{\mathrm{B}}=0.548$ K, $A/k_{\mathrm{B}}=1.1\times 10^{-3}$ K,
\cite{miretal99prl,barkenrumhencri03prl} and $\hat{H}^{\prime }$ is the part
of the Hamiltonian that does not commute with $S_{z},$ nonessential for $%
C_{S}(T)$ in Mn$_{12}.$ With the spin levels $\varepsilon
_{m}=-Dm^{2}-Am^{4} $ the energy is given by $U_{S}=\left( 1/Z\right)
\sum_{m=-S}^{S}\varepsilon _{m}\exp \left[ -\varepsilon _{m}/\left( k_{%
\mathrm{B}}T\right) \right] ,$ where $Z=\sum_{m=-S}^{S}\exp \left[
-\varepsilon _{m}/\left( k_{\mathrm{B}}T\right) \right] .$ Then $%
C_{S}=dU_{S}/dT,$ and fitting the lattice heat capacity $C_{\mathrm{ph}%
}=C-C_{S}$ yields the EDM curve shown in Fig.\ \ref{Fig-C} that is in an
excellent accord with the experimental data in the entire temperature range.

In fact, the lattice heat capacity also contains the contribution of nuclear
spins $\sim 1/T$ that is small in the Kelvin range, as well as the $T$%
-linear term that must be an artefact of the experimental procedure. To
eliminate the parasite $T$-term and visualize the $C_{\mathrm{ph}}\varpropto
T^{3}$ dependence at low temperatures, it is convenient to plot $C_{\mathrm{%
ph}}/T$ \ vs $T^{2}$ to get a straight line, as shown in Fig.\ \ref
{Fig-C_over_T}. The slope of the straight line yields the average value of
the Debye temperature $\bar{\Theta}_{\mathrm{D}}\equiv \left( \sum_{\lambda
}\Theta _{\mathrm{D},\lambda }^{-3}/3\right) ^{-1/3}=38$ K, in accord with
Ref.\ \onlinecite{gometal98prb}, whereas the coefficient in the $T$-linear
term is 0.022. Subtracting this small constant and fitting the rest, one
obtains
\begin{equation}
\Theta _{\mathrm{D},1}=29.6\ \text{K,\quad }\Theta _{\mathrm{D},2}=47.8\
\text{K,\quad }\Theta _{\mathrm{D},3}=61.0\ \text{K,}  \label{ThetaD123Res}
\end{equation}
as well as
\begin{equation}
\Theta _{\mathrm{G}}=15.6\ \text{K.}  \label{ThetaGRes}
\end{equation}
From Eqs.\ (\ref{OmegaDDef}) and (\ref{ThetaD123Res}) follows
\begin{equation}
v_{1}=1541\ \text{m/s,\quad }v_{2}=2488\ \text{m/s,\quad }v_{3}=3176\ \text{%
m/s.}  \label{v123Res}
\end{equation}
The first of these speeds of sound should correspond to a nearly transverse
mode, while the last one should correspond to a nearly longitudinal mode.
The former is the most important in relaxation.

Of course, one can say that with enough fitting parameters one can fit any
function. While in general it is true, the scheme used here is a minimal
model with no excessive fitting parameters. Accounting for optical phonons
with a single parameter $\Theta _{\mathrm{G}}$ is a must, and there are no
physical reasons to set speeds of sound the same. The experimental results
in the natural representation in Fig.\ \ref{Fig-C} can also be fitted by the
EDM with one acoustic phonon mode ($v_{1}=v_{2}=v_{3}=v$) and two acoustic
phonon modes ($v_{1}=v_{2}=v_{t},$ $v_{3}=v_{l}$), and the results are
visually not dramatically worse. However, this fitting method is inferior
since it puts more weight on the high-temperature range and tends to neglect
the low-temperature range. It is better to fit the $C/T$ data with the
subtraction of the parasite $T$-term, as was done above (see Fig.\ \ref
{Fig-C_over_T}). Then one can see more difference between different fits.
The most stringent check of different fits can be achieved in the most
balanced representation of $C(T)$ over the whole temperature range using $%
C/T^{2}$ (with subtracted $T$-term) as the fitting target. The results shown
in Fig.\ \ref{Fig-C_over_T2} demonstrate that only the model with three
different speeds of sound really fits the data. In addition, for this model
results obtained with different fitting methods do not differ much. Fitting
of $C/T^{2}$ with three phonon modes yields $\Theta _{\mathrm{D},\lambda
}=29.7,$ 47.1, 61.4 K and $\Theta _{\mathrm{G}}=15.7\ $K that is very close
to the results of the $C/T$ fitting, Eqs.\ (\ref{ThetaD123Res}) and (\ref
{ThetaGRes}), and even to the results of the $C$ fitting ($\Theta _{\mathrm{D%
},\lambda }=29.5,$ 50.0, 57.0 K and $\Theta _{\mathrm{G}}=15.8\ $K). To the
contrary, models with one or two different speeds of sound yield very
different results with the three different fitting schemes. Thus one
concludes that these models do not work. Still we quote the results from
fitting $C/T^{2}$ in Fig.\ \ref{Fig-C_over_T2} for the reference: One phonon
mode $\Theta _{\mathrm{D}}=39.4$ K and $\Theta _{\mathrm{G}}=18.4\ $K, two
phonon modes $\Theta _{\mathrm{D},t}=38.1$ K, $\Theta _{\mathrm{D},l}=41.3$
K, and $\Theta _{\mathrm{G}}=17.5\ $K.

\begin{figure}[t]
\unitlength1cm
\begin{picture}(11,5.5)
\centerline{\psfig{file=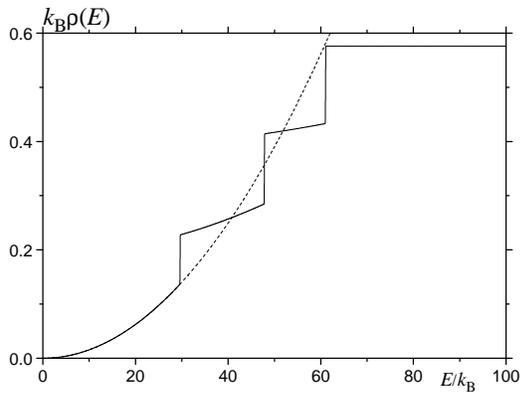,angle=-90,width=8cm}}
\end{picture}
\caption{Phonon density of states for Mn$_{12}$ within the extended Debye
model.}
\label{Fig-DOS}
\end{figure}

Fig.\ \ref{Fig-DOS} shows the density of states in Mn$_{12}$ within the
extended Debye model with parameters given by Eqs.\ (\ref{ThetaD123Res}) and
(\ref{ThetaGRes}). Although there are steps in the DOS that reflect the
crudeness of the underlying Debye model, there are three smaller steps
instead of a single large step in the original Debye model. Thus the EDM is
much more realistic than the DM. The accuracy of the extracted data on
speeds of sound is difficult to estimate because of the assumption of the
rigid cut-off at $k_{\mathrm{D}}.$ One rather should consider Eqs.\ (\ref
{ThetaD123Res})--(\ref{v123Res}) as qualitative results that capture some
physics of phonons in MM.

A practical question is how to apply the results obtained above to the
relaxation in MM. All existing formulas are based on the model with one
longitudinal phonon mode and two transverse phonon modes. Within the
mechanism of the molecule rotation without distortion, the contributions of
the processes contain the factors $\sin ^{2}\theta =\left[ \mathbf{k\times e}%
_{\mathbf{k}\lambda }\right] ^{2}/k^{2},$ where $\mathbf{e}_{\mathbf{k}%
\lambda }$ are phonon polarization vectors. Thus longitudinal phonons do not
make a contribution while transverse phonons make the maximal contribution.
In reality phonons are not purely longitudinal and purely transverse and it
is difficult to extract the angle between $\mathbf{k}$ and $\mathbf{e}_{%
\mathbf{k}\lambda }.$ For an estimation one can propose a rule of thumb: $%
\theta =\pi /2,$ $\pi /4,0$ for the phonon modes with $v_{1},$ $v_{2},$ and $%
v_{3},$ respectively. With this conjecture, one can make the following
replacement in the formulas for the rates of direct processes:
\begin{equation}
\frac{1}{v_{t}^{5}}\Rightarrow \frac{1}{2}\left( \frac{1}{v_{1}^{5}}+\frac{%
1/2}{v_{2}^{5}}\right) =\frac{1}{2v_{1}^{5}}\left[ 1+\frac{1}{2}\left( \frac{%
v_{1}}{v_{2}}\right) ^{5}\right] .  \label{vt5Replacement}
\end{equation}
For the values listed in Eq.\ (\ref{v123Res}), the correction term in square
brackets is only 0.046 and can be neglected. Thus the rule of thumb
simplifies to keeping only the softest mode with the coefficient 1/2. Then
the increase of the rate due to using the model with three different phonon
modes instead of the traditional model with one longitudinal and two
degenerate transverse phonon modes is given by $v_{t}^{5}/(2v_{1}^{5})=%
\Theta _{\mathrm{D},t}^{5}/(2\Theta _{\mathrm{D},1}^{5}).$ With $\Theta _{%
\mathrm{D},1}=29.7$ and $\Theta _{\mathrm{D},t}=38.1$ K obtained above, the
rate increase makes up $\Theta _{\mathrm{D},t}^{5}/(2\Theta _{\mathrm{D}%
,1}^{5})=2.4.$

The author thanks Marco Evangelisti for providing many illuminating insights
into the low-temperature heat capacity measurements, Jonathan Friedman for
stimulating this research, and Eugene Chudnovsky for useful discussions.
This work has been supported by the NSF Grant No. DMR-0703639.


\end{document}